# Extrinsic localized excitons in patterned 2D semiconductors

D Yagodkin[1], K Greben[1], A Eljarrat[2], S Kovalchuk[1], M Ghorbani-Asl[3], M Jain[3], S Kretschmer[3], N Severin[2], J P Rabe[2], A V Krasheninnikov[3,4], C T Koch[2], and K I Bolotin[1]

[1] *Department of Physics, Freie Universität Berlin, Berlin, Germany*
[2] *Department of Physics & IRIS Adlershof, Humboldt-Universität zu Berlin, Berlin, Germany*
[3] *Helmholtz-Zentrum Dresden-Rossendorf, Institute of Ion Beam Physics and Materials Research, Dresden, Germany*
[4] *Department of Applied Physics, Aalto University, Aalto, Finland*

We demonstrate a new localized excitonic state in patterned monolayer 2D semiconductors. This state is not associated with lattice disorder but is extrinsic, i.e. results from external molecules on the material surface. The signature of an exciton associated with that state is observed in the photoluminescence spectrum after electron beam exposure of several 2D semiconductors. The localized state, which is distinguished by non-linear power dependence, survives up to room temperature and is patternable down to 20 nm resolution. We probe the response of the new exciton to the changes of electron energy, nanomechanical cleaning, and encapsulation via multiple microscopic, spectroscopic, and computational techniques. All these approaches suggest that the state does not originate from irradiation-induced structural defects or spatially non-uniform strain, as commonly assumed. Instead, we show that it is extrinsic, likely a charge transfer exciton associated with the organic substance deposited onto the 2D semiconductor. By demonstrating that structural defects are not required for the formation of localized excitons, our work opens new possibilities for further understanding of these states and using them for example in chemical sensing and quantum technologies.

Excitonic complexes in transition metal dichalcogenides (TMDs) are Coulomb-bound states of electrons and holes, which can be broadly divided into free and localized. Free excitons move about the crystal transmitting information and energy[1]. The properties of these quasiparticles (e.g. neutral and charged, dark and bright free excitons), including their binding energies, formation mechanisms, and valley dynamics have largely been understood both theoretically and experimentally within the last decade[2,3]. In contrast, localized excitons are spatially confined by a sufficiently strong potential. Unlike their free counterparts, localized excitons do not transfer information or energy, couple to both valleys of the TMD, and have a nearly flat energy dispersion[4–6]. Although these states were previously largely ignored, many recent developments suggest that they may determine critical properties of TMDs, such as the diffusion length, the doping response, and lifetimes of various excitonic species[1,7]. Some types of localized excitons feature charge and spin lifetimes up to microseconds owing to the decoupling from the host crystal[8,9]. These states may play a determining role in the relaxation of excitons in TMD heterostructures by providing pathways for excitations to tunnel between two angle-mismatched TMDs[7,10]. Some localized states behave as single quantum emitters driving the interest in potential applications in quantum technologies[6,9,11,12]. Finally, long-lived localized states induced right at the surface of TMDs may potentially be used as sensors of a TMD's environment[13,14].

While many different types of localized excitons have been observed in different TMDs[7], the mechanisms leading to their formation as well as their intrinsic properties remain largely unknown. These states are near-universally attributed to structural disorder in the TMD lattice[15]. The prevalent source of such disorder is lattice vacancies that can be either native[4] or induced using high-energy particles[16,17]. Only very recently, combined local probe/spectroscopy studies of ultraclean encapsulated samples managed to pinpoint a certain localized state to a structural vacancy defect in TMDs[4,18]. Another type of structural disorder appears when spatially non-uniform strain locally distorts a TMD lattice[19]. Such a strain can lead to the formation of a localized state by creating a spatially varying potential confining the excitons. The nature of such a localized state and the role of structural vacancies in its formation remain debated[20].

Finally, the periodic disorder arising in angle-matched TMD heterostructures (moiré lattices) produces periodic potential wells deep enough to localize excitons[21–23].

Contrary to the examples discussed above, we demonstrate a controlled and patternable localized state in TMDs that is not related to the intrinsic disorder. We induce a localized excitonic state in a broad family of TMD materials via electron irradiation and show that multiple key characteristics of this state – its changes with the energy of the electrons, the response to nanomechanical cleaning, and the response to hBN encapsulation is not compatible with the structural disorder, such as lattice defects or localized strain. We suggest that the only mechanism that can explain such a state is the formation of a charge-transfer exciton between an organic molecule on the TMD surface and the TMD itself. Our work may potentially clarify the nature of some previously observed localized states and provide avenues to patterning and tuning such states in the future.

From the family of TMD materials, we focus mainly on $MoS_2$ while also considering $WSe_2$ and $WS_2$ samples. Native defects and related excitonic states manifested in the photoluminescence (PL) spectrum of these materials are well charachterized[4,15,24,25]. A monolayer flake is mechanically exfoliated and transferred onto a $SiO_2$ or hBN substrate (Methods), loaded into the electron beam lithography (EBL) system where an electron beam is rastered across the TMD surface. Unless stated otherwise, the acceleration voltage of the electron beam is 10 keV and the dose (fluence of electrons) is varied in the range 0.5 – 14 $mC/cm^2$ (Supplementary Note 1). We raster the beam in a pattern consisting of stripes with a spacing of 80 nm (Fig. 1a, Methods). This pattern is visible in the atomic force microscopy (AFM) topography images (Fig. 1b) of the TMD's surface after exposure without any additional treatment (details in Supplementary Note 1). The exposed regions are about 0.3 nm higher (bottom panel in Fig. 1b) than the unexposed regions. The height profile has a period of ~80 nm, consistent with the designed e-beam pattern. A similar pattern is also observed using scanning transmission electron microscopy (STEM) of suspended bilayer $MoS_2$ exposed to a higher electron dose of 14 $mC/cm^2$ (Fig. 1c). A single stripe has a width of ~20 nm and the period between stripes is ~80 nm (bottom panel in Fig. 1c). The appearance of an extra layer of material after exposure to e-beam is known and signals e-beam assisted deposition of molecules onto the surface of $MoS_2$[26–28].

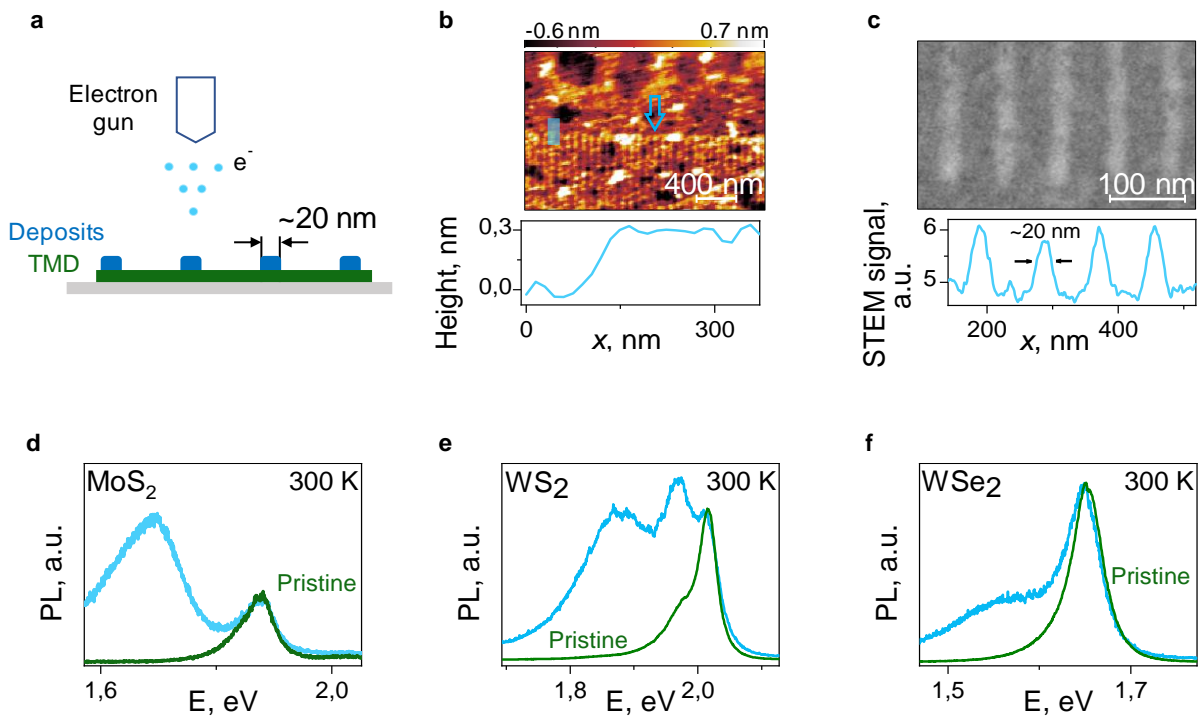

**Fig. 1: EBL Nanopatterning. a)** Local e-beam exposure of TMDs to generate a new excitonic state. **b)** AFM topography of a 1L-$MoS_2$ on hBN sample (D1) after e-beam patterning. The period of the lines seen in the image, 80 nm, matches the EBL pattern (vertical lines, cyan arrow). The height profile along the blue line in the image is shown in the bottom panel (linear background was subtracted). **c)** STEM image of a suspended $MoS_2$ sample (D2) after EBL along with a line profile (bottom panel). The deposits in the e-beam treated regions of the suspended bilayer $MoS_2$ are also visible. The width of the stripes, about 20 nm (bottom panel), and their period, 80 nm, match the parameters of EBL patterning. **d-f)** Normalized PL spectra from pristine (green curve) and exposed to e-beam (blue curve) areas of $MoS_2$, $WS_2$, and $WSe_2$ correspondingly.

In the region of the $MoS_2$ sample which was exposed to electrons we observe a new peak in the PL spectrum at around 1.7 eV. The feature is seen even at room temperature under ambient conditions (blue curve in Fig. 1d). It is absent in unexposed areas of the same device before patterning (green curve in Fig. 1d), as well as in other pristine $MoS_2$ samples or on e-beam treated bare substrate. The peak is well-separated from native neutral (1.86 eV) and charged (1.83 eV) excitons in $MoS_2$[29,30]. A similar peak also appears in other TMDs ($WS_2$, $WSe_2$) after electron exposure with a similar ~150 meV shift relative to the neutral (A) exciton (Fig. 1e, f). We tentatively label this new feature "E peak" and turn to the investigation of its properties and origin.

We start by studying the low-temperature PL spectra of the patterned $MoS_2$ device. Figure 2a shows PL spectra of the sample at 15 K under an excitation power of 50 nW – 2.8 mW (Methods). The spectrum at high fluence consists of a 40 meV wide E peak centered at 1.735 eV and a shoulder of another peak at 1.77 eV (Supplementary Fig. S1). The latter peak is commonly attributed to native defect-bound excitons in $MoS_2$[15,24,25]. At lower fluence, emission of all other excitons is suppressed (Fig. 2 and S2) while sharp features withing the E peak emerge (vertical dashed lines). The FWHM of these features of around 1 meV is much smaller compared to the FWHM of free excitons of 8 meV (Fig. S1). The E peak exhibits a sub-linear power dependence ($I \sim P^{0.38}$), compared to the linear dependence of free excitons (Fig. 2a, inset). This is a tell-tale sign of a localized state[4,31,32]. The overall spectral shape is likely a combination of multiple narrow peak from individual states distributed between 1.70 – 1.74 eV due to the varying dielectric background[14,33]. Localized emitters with similar power dependence have been observed in various 2D materials and ascribed to structural disorder[4,7,32,34,35]. Notably, while previously reported defect-related excitonic peaks in TMDs are only observed at cryogenic temperature[15,24,25,36,37], the E peak persists up to room temperature (Fig. 1d-f).

Next, we study the dependence of the state on the density of free carriers. We apply a gate voltage ($V_g$) between the Si backgate and the 1L-$MoS_2$. We observe that the E peak intensity is strongly $V_g$-dependent with the maximum observed at negative $V_g$ (Fig. 2b). The E peak is absent in the pristine region at any voltage (Fig. S3). By examining the intensities of neutral (A) and negatively-charged (T) excitonic peaks at 1.93 and 1.90 eV respectively, we establish the E peak is visible in the range of gate voltages corresponding to the Fermi level being close to the bottom of the conduction band of $MoS_2$[15]. Such a gate voltage dependence has also been observed for other types of localized excitons[4]. It suggests that the E peak involves a shallow donor state inside the bandgap of $MoS_2$ close to the conduction band which is filled at positive gate voltage[15]. The occupancy of that defect state changes within the range of applied gate voltage producing the observed variation in the associated exciton emission[38,39]. The electrical tunability of this state provides a simple tool for its control[40].

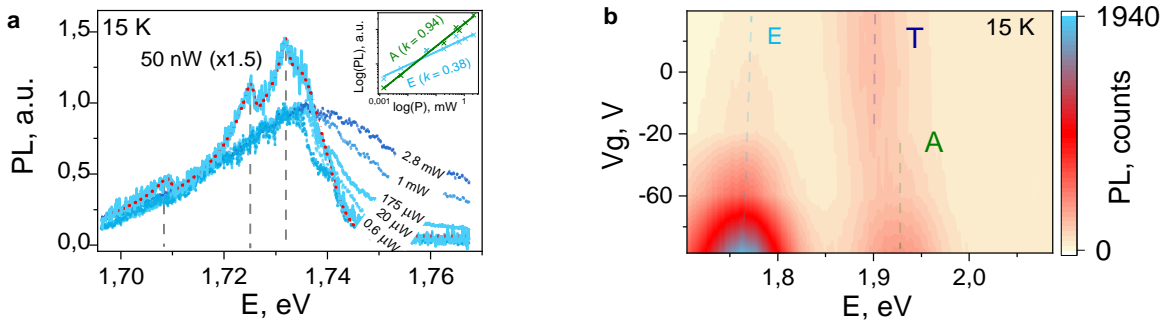

**Fig. 2: E-beam-related photoluminescence peak and its fluence- and gate-dependencies. a)** Normalized PL spectrum of e-beam exposed and later hBN encapsulated 1L-$MoS_2$ (sample D3) at 15 K, measured at excitation powers ranging from 50 nW (solid blue curve, multiplied by x1.5 for better visibility) to 2.8 mW. Inset: Power dependence of the area under the E- and A-peak in PL (blue and green crosses respectively). The E-peak has a sublinear dependence on power (blue line is a linear fit with a slope of 0.38) which is expected for a localized state. **b)** Low temperature PL spectra of the modified $MoS_2$ flake on $SiO_2$/Si substrate (sample D4) at back gate voltages from -80 V to 20 V.

Having established that the E peak seen in Figures 1 and 2 is caused by a localized state, we now investigate its microscopic origin. The most straightforward explanation would be that the state relates to the structural defect produced by e-beam exposure[41–43]. To test this hypothesis, we vary the electron energy from 1.6 keV to 30 keV while keeping the exposure dose constant and examine the resulting changes in the PL spectra. We interpret the ratio of the integrated areas under the E- and A-peak as proportional to the density of localized excitons (Fig. S5). We observe a strong electron energy dependence of that inferred quantity (points in Fig. 3a). The peak ratio (and hence the density of localized excitons) remains non-zero down to low electron energies (<5 keV) and decreases with energies above 10 keV.

The electron energy dependence of the inferred density of localized excitons is different compared to what is expected from structural defects induced by electron irradiation in 2D TMDs. In general, the formation of such defects by a combination of electronic excitations and knock-on damage requires electrons with an energy of at least 30 keV, and the density of structural defects should increase with the electron energy for the entire range of energies[43,44]. The presence of a sharp peak in Fig. 3a contradicts this scenario. These defects can also appear at lower (~ 100 eV) energies due to chemical etching resulting from radicals produced by the beam through electronic excitations[45], but this mechanism is minor in the used electron energy range.

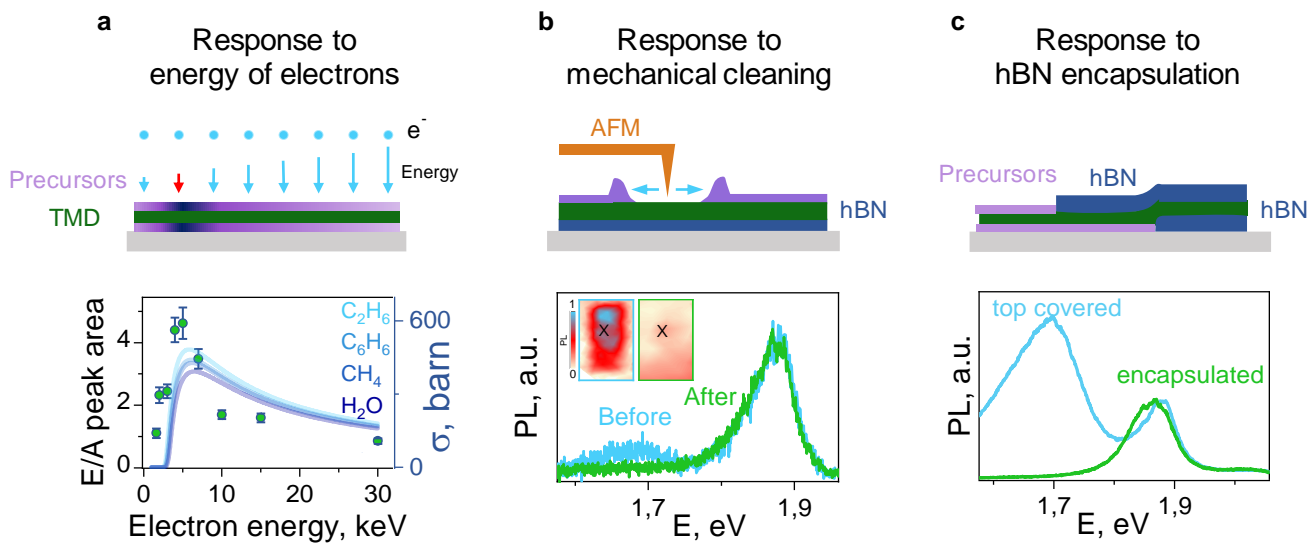

**Fig. 3: Nature of localized states. a)** Ratio of the areas under the E- and A-peak in PL (proportional to the density of induced localized states) vs. electron energy in the e-beam exposed area of the $MoS_2$ sample D5 (green dots, left axis). The electron dose was kept constant. Calculated cross section of electron scissoring of hydrogen in $C_2H_6$; $C_6H_6$; $CH_4$; $H_2O$ through ballistic displacement of hydrogen atoms (shades of blue, right axis). **b)** PL spectra of a low dose e-beam treated region of $MoS_2$ on hBN (sample D6) before and after mechanical cleaning with AFM (blue and green curves correspondingly). Inset: spatial maps of integrated E-peak at room temperature before (left) and after (right) AFM cleaning. The PL spectra are taken at the same point marked "X" before and after mechanical cleaning. While the E-peak is prominent before cleaning, it completely disappears after **c)** PL spectra of e-beam treated region of $MoS_2$ (sample D7) top-covered with hBN (light blue), and fully encapsulated (green). The E peak appears in either top- or bottom hBN (Fig. 3c, 3b) covered samples, but not in fully encapsulated devices.

The hypothesis that the E-peak is unrelated to TMD structural defects is further supported by two additional experiments. In the first experiment, the $MoS_2$ flake on hBN (for mechanical support) is exposed to a low electron dose (Supplementary Note 1) in a rectangular area (inset in Fig. 3b) and the PL spectrum was recorded (Fig. 3b, blue curve). Then, we nanomechanically clean the surface of the TMD using repeated scanning with an AFM tip (the "nano-squeegee" technique[46]) and examine the resulting changes in the PL spectra. To achieve this, we imaged the entire device in an AFM while gradually increasing the contact force up to 300 nN over 12 hours. As expected[46], this procedure squeezed the deposits produced by the e-beam out of the AFM imaging window. After the procedure, the E peak in the imaged region disappears completely (Fig. 3b, green curve). This would not be expected if structural defects were the origin of the E peak. In the second experiment, we used a sample that contains area of $MoS_2$ on $SiO_2$ covered with hBN only on top as well as an $MoS_2$ flake area encapsulated in hBN. After electron exposure, the $MoS_2$ on $SiO_2$ shows the high intensity of the E-peak suggesting that electrons penetrate the thin (8 nm) hBN layer (Fig. 3c). In contrast, the E peak does not arise in the region of the same flake encapsulated in hBN. Again, this behavior is not consistent with structural defects whose density should not depend on the presence of a thin bottom hBN layer. We exclude hBN-related change of optical properties or dielectric environment[2,14] as if an $MoS_2$ flake on hBN is first exposed to e-beam and then covered with top hBN, the E peak is observed (Fig. 2a, S2).

The arguments above prove that the E-peak is not related to lattice defects in the TMD. Other mechanisms commonly invoked to explain the appearance of localized states such as strain-, crystalline phase-, or dielectric- related localization can also be excluded in our case. First, the E-peak-related state does not result from the localization of a neutral exciton due to the varying dielectric environment of $MoS_2$. The observed energy separation between the neutral exciton and the E peak of 195 meV (Fig. S1, 2a) is much larger than the maximum energy shift of 50 meV which can be induced by changes in the dielectric environment, 50 meV[14,47]. Second, the state cannot be explained by the localization of excitons in a spatially inhomogeneous strain field. While such strain can, in principle, localize the excitons[15,22,39,48–50], the maximum strain estimated from AFM topography is an order of magnitude smaller than what is required for this mechanism to explain our data[51]. Third, local patterning can result in a structural phase transition (2H → 1T) breaking the material into individual quantum dots[52,53], thereby localizing the excitons. Our STEM imaging (Fig. S5) confirms the uniformity of the crystal lattice across pristine and e-beam exposed regions, contradicting such an argument.

We believe that the only plausible microscopic mechanism behind the localized state consistent with our data is that it relates to organic molecules on the surface of $MoS_2$. Electron beam irradiation is known to both break and modify organic residues already present on the sample surface[26–28]. Such deposits are observed both in our AFM and STEM images (Fig. 1a,b). The disappearance of the E-peak after nanomechanical cleaning (Fig. 3b) is consistent with these molecules being squeezed out from the scanning area. The absence of the E-peak in the encapsulated device (Fig. 3c) is also consistent with the pristine molecule-free nature of the hBN/TMD/hBN devices.

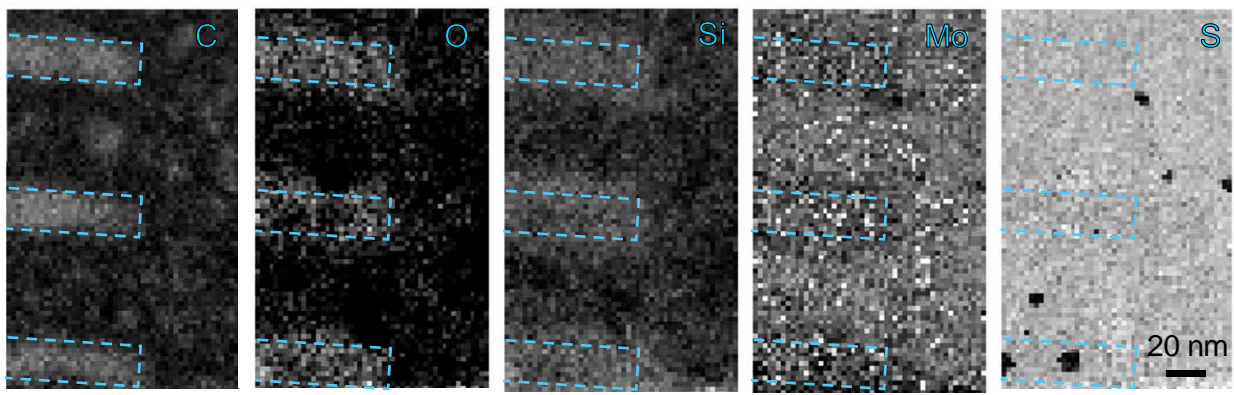

**Fig. 4:** Spatial maps of EELS signal for C, O, Si, Mo, and S (white – more intense) across the e-beam patterned region, sample D2 (blue dashed line shows region boundaries). While EELS signal for Mo and S is uniform across the entire specimen, it surges in the patterned regions for C, O, and Si. Thus, we attributed deposits to organics.

What is the chemical nature of the organic molecules on the surface of the TMD and how do these molecules influence the E peak formation? We hypothesize that the electron beam radicalizes molecular residues present on the surface by ballistic displacement of H atoms (based on the low energy of the peak in Fig. 3a). Using density functional theory (DFT)-based molecular dynamics (MD), we calculate the displacement thresholds, which correspond to the minimum kinetic energy needed to permanently displace an H atom. Then, we apply the McKinley–Feshbach formalism to calculate the cross-section of that process (Methods) and compare it with the experimental data. For a range of organic hydrogen-containing molecules ($C_2H_6$, $C_6H_6$, $CH_4$, and $H_2O$), we observe a dependence of the simulated cross-section on the electron energy (curves in Fig. 3a) that is very similar to the experimentally observed energy dependence of the localized state density inferred from the E- to A-peak ratio (points in Fig. 3a). Next, we run DFT electronic-structure calculation to obtain a modified density of states of $MoS_2$ in presence of the resulting molecules and single atoms. Both carbon and silicon adatoms produce states at the conduction band bottom (0.18 and 0.24 eV from the bottom correspondingly) while oxygen has little influence on the DOS (Fig. S6). Qualitatively similar signatures are seen for radicals containing C, Si, H, such as $CH_2$, $CH_3$, $C_6H_5$, and SiH (Fig. S7). The position of the states in the gap obtained from the calculations is consistent with the experimentally observed energy separation of E and A peaks (Fig. 2a, S1) and the response of PL to electrical doping (Fig. 2b). The presence of such organic molecules is furthermore confirmed by high-resolution electron energy loss spectroscopy (EELS) across the patterned region of $MoS_2$ on a TEM grid (Fig. 4). It shows a fivefold increase of the oxygen and carbon content and a tripled amount of silicon in e-beam treated areas (Fig. S8 and S9). Maps of the Mo and S signals (Fig. 4) do not show any variation between patterned and not patterned regions beyond additional scattering on Si (Fig. S9a), confirming the absence of e-beam induced structural defects. Again, these signatures point towards organic molecules interacting with the sample.

Overall, all our data suggests that the E peak is related to carbon- and silicon- containing radicals formed after ballistic displacement of hydrogen in residues on the sample surface. One possible scenario is that the E peak is a charge-transfer exciton between the TMD and the state provided by the organic radicals (Fig. S6, S7). Indeed, such excitons have previously been observed between TMDs and organic molecules on their surfaces[54,55]. These excitons are expected to have properties similar to that of the E peak: They only appear in the presence of and change with the concentration of external material (Fig. S5)[56,57], blueshift at high excitation powers (Fig. 2a and Fig. S2b)[58], and become a ground state at low temperature (Fig. 2a, S1). The state is expected to appear in a comparable spectral range in all TMD materials.

To conclude, we patterned bright, room-temperature, localized excitons in 2D semiconductors using conventional electron beam lithography. This easy-to-use method features nanoscale resolution along with the potential of wide tunability and scalability. We investigated the nature of the localized state using a combination of techniques and concluded that it is not related to structural defects in the TMD. Instead, the state emerges due to the ballistic displacement of hydrogen atoms from molecules present on the surface of the sample. The resulting organic radicals bonding to the TMD give rise to an in-bandgap state. The E-peak-related state could be a charge transfer exciton formed between e.g., a hole in the TMD and an electron in such a state provided by the organic radical. The observation of this state opens many interesting possibilities. First, its expected out-of-plane orientation should result in a large Stark shift in an out-of-plane electric field, further extending its spectral tunability. In the future, it will be interesting to measure this shift to establish the size of the state. Second, it is important to establish whether some of the localized excitons reported previously are related to such external molecules or indeed originate from structural defects. The two mechanisms can be easily distinguished by comparing photoluminescence spectra at low fluences before and after nanomechanical cleaning. Third, it will be interesting to controllably pattern closely spaced deterministic arrays of electrically controlled localized excitons and study the effects caused by exciton interactions. Finally, as the observed state originates from molecules localized right at the TMD's surface, we expect a stronger response to the dielectric screening compared to that of free excitons. Such a response combined with the observed room-temperature photoluminescence and environmental stability may be interesting for biological and chemical sensing.

**Methods:**

**Fabrication:** The $MoS_2$ on $SiO_2$ samples were obtained by mechanical exfoliation of bulk material (>99.9% pure synthetic crystals from HQ graphene) onto the 300 nm $SiO_2$/Si using scotch tape or PDMS methods. The samples were washed with acetone and IPA to remove the organic residues. Then, they are loaded into a Raith Pioneer II SEM/EBL for patterning. We used the dose and acceleration voltage specified in the main text with a beam current of 0.22 nA and 30 µm apertures. Patterning was done in the dot-by-dot regime with an 18.9 nm step between dots in a row and 80 nm step between rows. Afterwards, the presence of the E peak in the room temperature PL spectrum of the sample confirms successful modification. Finally, the samples are annealed in vacuum at 230 °C for about 12 h.

**Photoluminescence spectroscopy:** The dose dependence and E peak maps (Fig. 2c and 3c) were measured with a XploRA™ HORIBA using 532 nm excitation at 16 µW focused into a diffraction limited spot (~1 µm diameter). Low-temperature experiments (Fig. 2a and 2b) were performed with Witec Alpha confocal spectroscopy setup in an optical cryostat with 532 nm green laser excitation focused in a ~0.6 µm diameter spot. Low power PL spectra were fitted with four exponentially modified Gaussians to account for phononic sidebands.

**AFM:** Mechanical cleaning (nanosqueegee) and topography scans were performed with a NanoWizard® AFM in ambient conditions. For nanosqueegee we used non-contact tips (Tap300Al-G from BudgetSensors) in contact regime applying above 100 nN and contact tips (CONTPt-10 from NanoWorld) below 100 nN. The topography scan (Fig. 1a) was taken in contact mode with a 2 nN setpoint, two consecutive scans were always recorded. Afterward, polynomial background subtraction and line matching were applied to the scan.

**Theoretical results:** DFT calculations were performed using the Vienna ab-initio Simulation Package (VASP)[59,60] using the Perdew–Burke–Ernzerhof (PBE) exchange-correlation functional[61]. The displacement calculations with account for thermal vibrations were carried out using the McKinley-Feshbach formalism as discussed previously[62]. Further details of calculations can be found in the Supplementary Note 2.

**STEM and EELS:** The 2L – $MoS_2$ sample was mechanically exfoliated on PDMS and transferred onto a Quantifoil grid. Prior to its insertion into the electron microscope, the sample was baked for several hours at an elevated temperature (~130°C) in high vacuum (~$2\times10^{-6}$ Torr). STEM-EELS experiments are carried out using a Nikon HERMES microscope. This instrument is equipped with an aberration corrector, a cold-field-emission-gun (cFEG), a monochromator at ground potential, and a hybrid-pixel direct-detection camera (Dectris ELA). For the spectra presented in the main text, an energy dispersion of 0.8 eV/channel was used and the acceleration voltage was 60 keV. The probe convergence and EEL spectrometer aperture semi-angles were both around 36 mrad. To avoid the amorphous carbon film of the TEM grid the spectra are acquired at the freely suspended region, several tens of nanometers away from the film. Model based quantification of the spectra was performed after plural scattering deconvolution, using multi-linear least square (MLLS) fitting. The fitted model contains a power-law background function and Hartree-Slater generalized oscillator strength (GOS) functions for the edges; carbon and oxygen K, silicon and sulfur L3-1 and molybdenum M5-1. In this manner, absolute values for atomic ratios are obtained for the above-mentioned species[63,64].

Acknowledgments

We thank Nele Stetzuhn Abhijeet Kumar for great comments on the paper. We acknowledge the German Research Foundation (DFG) for financial support through the Collaborative Research Center TRR 227 Ultrafast Spin Dynamics (project B08), SfB 951 (projects A6, A12, B15, Z2). AVK further thanks DFG (projects KR 4866/2-1 and SFB-1415-417590517) for support. We thank HRLS Stuttgart, Germany and TU Dresden (Taurus cluster) for generously grants of computing time.

Competing Interests

The authors declare that there are no competing interests.

Author Contribution

D.Y., K.G., and K.I.B. conceived and designed the experiments, D.Y., K.G. prepared the samples, A.E. and C.T.K. performed EELS, STEM measurments along with related modeling and analysis, D.Y., N.S., and J.P.R. performed AFM measurements and AFM mechanical cleaning, D.Y., K.G. and S.K. performed the optical measurements, D.Y. analyzed the optical and AFM data, M. J., S. K., M. G. and A. K. performed the calculations and help to rationalize the experimental data, D.Y. and K.I.B. wrote the manuscript with input from all coauthors.

Data Availability

The data that support the findings of this study are available from the corresponding author upon reasonable request.

1. Mueller, T. & Malic, E. Exciton physics and device application of two-dimensional transition metal dichalcogenide semiconductors. *npj 2D Mater. Appl.* **2**, 29 (2018).
2. Wang, G. *et al.* Colloquium: Excitons in atomically thin transition metal dichalcogenides. (2018) doi:10.1103/RevModPhys.90.021001.
3. Xu, X., Yao, W., Xiao, D. & Heinz, T. F. Spin and pseudospins in layered transition metal dichalcogenides. (2014) doi:10.1038/NPHYS2942.
4. Rivera, P. *et al.* Intrinsic donor-bound excitons in ultraclean monolayer semiconductors. *Nat. Commun.* **12**, (2021).
5. Zheng, Y. J. *et al.* Point Defects and Localized Excitons in 2D WSe2. *ACS Nano* **13**, 6050–6059 (2019).
6. Li, Z. *et al.* Interlayer Exciton Transport in MoSe 2 /WSe 2 Heterostructures. (2021) doi:10.1021/acsnano.0c08981.
7. Chakraborty, C., Vamivakas, N. & Englund, D. Advances in quantum light emission from 2D materials. *Nanophotonics* **8**, 2017–2032 (2019).
8. Srivastava, A. *et al.* Optically active quantum dots in monolayer WSe 2. *Nat. Nanotechnol.* **10**, 491–496 (2015).
9. Moody, G. *et al.* Microsecond Valley Lifetime of Defect-Bound Excitons in Monolayer

10. Kumar, A. *et al.* Spin/Valley Coupled Dynamics of Electrons and Holes at the MoS 2 – MoSe 2 Interface. **15**, 22 (2022).
11. He, Y.-M. *et al.* Cascaded emission of single photons from the biexciton in monolayered WSe2. *Nat. Commun.* **7**, 13409 (2016).
12. Dass, C. K. *et al.* Ultra-Long Lifetimes of Single Quantum Emitters in Monolayer WSe 2 /hBN Heterostructures . *Adv. Quantum Technol.* **2**, 1900022 (2019).
13. Yu, X. *et al.* Graphene-based smart materials. *Nat. Rev. Mater.* **2**, 1–14 (2017).
14. Raja, A. *et al.* Dielectric disorder in two-dimensional materials. *Nat. Nanotechnol.* **14**, 832–837 (2019).
15. Greben, K., Arora, S., Harats, M. G. & Bolotin, K. I. Intrinsic and Extrinsic Defect-Related Excitons in TMDCs. *Nano Lett.* **20**, 2544–2550 (2020).
16. Klein, J. *et al.* Robust valley polarization of helium ion modified atomically thin MoS2. *2D Mater.* **5**, (2018).
17. Klein, J. *et al.* Site-selectively generated photon emitters in monolayer MoS2 via local helium ion irradiation. *Nat. Commun.* **10**, 2755 (2019).
18. Mitterreiter, E. *et al.* Atomistic Positioning of Defects in Helium Ion Treated Single-Layer MoS2. *Nano Lett.* **20**, 4437–4444 (2020).
19. Branny, A., Kumar, S., Proux, R. & Gerardot, B. D. Deterministic strain-induced arrays of quantum emitters in a two-dimensional semiconductor. *Nat. Commun.* **8**, 1–7 (2017).
20. Linhart, L. *et al.* Localized Intervalley Defect Excitons as Single-Photon Emitters in WSe 2. (2019) doi:10.1103/PhysRevLett.123.146401.
21. Wilson, N. P., Yao, W., Shan, J. & Xu, X. Excitons and emergent quantum phenomena in stacked 2D semiconductors. *Nat. 2021 5997885* **599**, 383–392 (2021).
22. Seyler, K. L. *et al.* Signatures of moiré-trapped valley excitons in MoSe2/WSe2 heterobilayers. *Nature* **567**, 66–70 (2019).
23. Wang, X. *et al.* Moiré trions in MoSe2/WSe2 heterobilayers. *Nat. Nanotechnol.* (2021) doi:10.1038/s41565-021-00969-2.
24. Cadiz, F. *et al.* Ultra-low power threshold for laser induced changes in optical properties of 2D molybdenum dichalcogenides. *2D Mater.* **3**, (2016).
25. Korn, T., Heydrich, S., Hirmer, M., Schmutzler, J. & Schller, C. Low-temperature photocarrier dynamics in monolayer MoS2. *Appl. Phys. Lett.* **99**, 102109 (2011).
26. Meyer, J. C., Girit, C. O., Crommie, M. F. & Zettl, A. Hydrocarbon lithography on graphene membranes. *Appl. Phys. Lett.* **92**, 123110 (2008).
27. Shen, X., Wang, H. & Yu, T. How do the electron beam writing and metal deposition affect the properties of graphene during device fabrication? *Nanoscale* **5**, 3352–3358 (2013).
28. Love, G., Scott, V. D., Dennis, N. M. T. & Laurenson, L. Sources of contamination in electron optical equipment. *Scanning* **4**, 32–39 (1981).
29. Mak, K. F. *et al.* Tightly bound trions in monolayer MoS 2. *Nat. Mater.* **12**, 207–211 (2013).
30. Mouri, S., Miyauchi, Y. & Matsuda, K. Tunable photoluminescence of monolayer MoS2 via chemical doping. *Nano Lett.* **13**, 5944–5948 (2013).
31. Schmidt, T., Lischka, K. & Zulehner, W. Excitation-power dependence of the near-band-edge photoluminescence of semiconductors. *Phys. Rev. B* **45**, 8989–8994 (1992).
32. Klein, J. *et al.* Atomistic defect states as quantum emitters in monolayer MoS2. *arXiv* **10**, 2755 (2019).
33. Klots, A. R. *et al.* Controlled dynamic screening of excitonic complexes in 2D semiconductors. *Sci. Reports 2018 81* **8**, 1–8 (2018).
34. Fournier, C. *et al.* Position-controlled quantum emitters with reproducible emission wavelength in hexagonal boron nitride. doi:10.1038/s41467-021-24019-6.
35. Tran, T. T., Bray, K., Ford, M. J., Toth, M. & Aharonovich, I. Quantum emission from hexagonal boron nitride monolayers. *Nat. Nanotechnol.* **11**, 37–41 (2016).

36. Klein, J. *et al.* Site-selectively generated photon emitters in monolayer MoS2 via local helium ion irradiation. *Nat. Commun.* **10**, 2755 (2019).
37. Klein, J. *et al.* Robust valley polarization of helium ion modified atomically thin MoS2. *2D Mater.* **5**, (2018).
38. Jauregui, L. A. *et al.* Electrical control of interlayer exciton dynamics in atomically thin heterostructures. *Science (80-. ).* **366**, 870–875 (2019).
39. Ciarrocchi, A. *et al.* Polarization switching and electrical control of interlayer excitons in two-dimensional van der Waals heterostructures. *Nat. Photonics* **13**, 131–136 (2019).
40. Hötger, A. *et al.* Gate-Switchable Arrays of Quantum Light Emitters in Contacted Monolayer MoS2 van der Waals Heterodevices. *Nano Lett.* **21**, 1040–1046 (2021).
41. Komsa, H. P. *et al.* Two-dimensional transition metal dichalcogenides under electron irradiation: Defect production and doping. *Phys. Rev. Lett.* **109**, (2012).
42. Komsa, H. P., Kurasch, S., Lehtinen, O., Kaiser, U. & Krasheninnikov, A. V. From point to extended defects in two-dimensional MoS2: Evolution of atomic structure under electron irradiation. *Phys. Rev. B - Condens. Matter Mater. Phys.* **88**, 35301 (2013).
43. Kretschmer, S., Lehnert, T., Kaiser, U. & Krasheninnikov, A. V. Formation of Defects in Two-Dimensional MoS2 in the Transmission Electron Microscope at Electron Energies below the Knock-on Threshold: The Role of Electronic Excitations. *Nano Lett.* **20**, 2865–2870 (2020).
44. Komsa, H. P. *et al.* Two-dimensional transition metal dichalcogenides under electron irradiation: Defect production and doping. *Phys. Rev. Lett.* **109**, 1–5 (2012).
45. Palmer, R. E. & Rous, P. J. Resonances in electron scattering by molecules on surfaces. *Rev. Mod. Phys.* **64**, 383 (1992).
46. Rosenberger, M. R. *et al.* Nano-"Squeegee" for the Creation of Clean 2D Material Interfaces. *ACS Appl. Mater. Interfaces* **10**, 10379–10387 (2018).
47. Lin, Y. *et al.* Dielectric screening of excitons and trions in single-layer MoS2. *Nano Lett.* **14**, 5569–5576 (2014).
48. Branny, A., Kumar, S., Proux, R. & Gerardot, B. D. Deterministic strain-induced arrays of quantum emitters in a two-dimensional semiconductor. *Nat. Commun.* **8**, (2017).
49. Darlington, T. P. *et al.* Imaging strain-localized excitons in nanoscale bubbles of monolayer WSe2 at room temperature. *Nat. Nanotechnol.* **15**, 854–860 (2020).
50. Kremser, M. *et al.* Discrete interactions between a few interlayer excitons trapped at a MoSe2–WSe2 heterointerface. *npj 2D Mater. Appl.* **4**, (2020).
51. Tyurnina, A. V. *et al.* Strained Bubbles in van der Waals Heterostructures as Local Emitters of Photoluminescence with Adjustable Wavelength. *ACS Photonics* **6**, 516–524 (2019).
52. Lin, Y. C., Dumcenco, D. O., Huang, Y. S. & Suenaga, K. Atomic mechanism of the semiconducting-to-metallic phase transition in single-layered MoS 2. *Nat. Nanotechnol.* **9**, 391–396 (2014).
53. Han, S. W. *et al.* Electron beam-formed ferromagnetic defects on MoS2 surface along 1 T phase transition. *Sci. Rep.* **6**, (2016).
54. Zhu, T. *et al.* Highly mobile charge-transfer excitons in two-dimensional WS2/tetracene heterostructures. *Sci. Adv.* **4**, 1–9 (2018).
55. Kafle, T. R. *et al.* Charge Transfer Exciton and Spin Flipping at Organic-Transition-Metal Dichalcogenide Interfaces. *ACS Nano* **11**, 10184–10192 (2017).
56. Lunz, M. *et al.* Influence of quantum dot concentration on Förster resonant energy transfer in monodispersed nanocrystal quantum dot monolayers. *Phys. Rev. B - Condens. Matter Mater. Phys.* **81**, (2010).
57. Kagan, C., Murray, C. & Bawendi, M. Long-range resonance transfer of electronic excitations in close-packed CdSe quantum-dot solids. *Phys. Rev. B - Condens. Matter Mater. Phys.* **54**, 8633–8643 (1996).
58. Nagler, P. *et al.* Interlayer exciton dynamics in a dichalcogenide monolayer heterostructure. *2D Mater.* **4**, (2017).

59. Kresse, G. & Joubert, D. From ultrasoft pseudopotentials to the projector augmented-wave method. *Phys. Rev. B* **59**, 1758 (1999).
60. Kresse, G. & Furthmüller, J. Efficient iterative schemes for *ab initio* total-energy calculations using a plane-wave basis set. *Phys. Rev. B* **54**, 11169 (1996).
61. Perdew, J. P., Burke, K. & Ernzerhof, M. Generalized Gradient Approximation Made Simple. *Phys. Rev. Lett.* **77**, 3865 (1996).
62. Meyer, J. C. *et al.* Accurate measurement of electron beam induced displacement cross sections for single-layer graphene. *Phys. Rev. Lett.* **108**, 196102 (2012).
63. Egerton, R. F. *Electron Energy-Loss Spectroscopy in the Electron Microscope*. *Electron Energy-Loss Spectroscopy in the Electron Microscope* (Springer US, 2011). doi:10.1007/978-1-4419-9583-4.
64. Peña, F. de la *et al.* hyperspy/hyperspy: Release v1.6.1. (2020) doi:10.5281/ZENODO.4294676.